\documentstyle[aps,prl,twocolumn,amssymb]{revtex}
\def\Z{\Bbb{Z}}
\def\C{\Bbb{C}}
\def\ad{{\,\rm ad}}
\def\tr{{\,\rm tr}}

\newcommand{\ket}[1]{|#1\rangle}

\begin{document}
\title{A new class of completely integrable quantum spin chains}
\author{Toma\v z Prosen}
\address{Physics Department, Faculty of Mathematics and Physics,
University of Ljubljana, Jadranska 19, 1111 Ljubljana, Slovenia}
\date{\today}
\draft
\maketitle
\begin{abstract}
\noindent
A large (infinitely-dimensional) class of completely integrable
(possibly non-autonomous) spin chains is discovered associated
to an infinite-dimensional Lie Algebra of infinite rank.
The complete set of integrals of motion is constructed explicitly,
as well as their eigenstates and spectra. As an example
we outline {\em kicked Ising model}: Ising chain periodically
kicked with transversal magnetic field.
\end{abstract}


\noindent
In the past three decades intricate algebraic techniques (under the names
{\em quantum inverse scattering} or {\em algebraic Bethe Ansatz})
have been developed \cite{KBI93} in order to construct Integrable
Quantum Many-body (IQM) dynamical systems and the associated complete sets
of integrals of motion. Integrability of a quantum many-body dynamical
systems is defined in a generalized Liouvilean sense; namely by
the existence of an infinite set of (independent and local)
conservation laws. All the so far discovered IQM systems are 
1-dimensional, typically SU(2) spin chains or related
systems. Quantum integrability is {\em nongeneric} but, however, of great
importance since it has been shown recently \cite{Prelovsek} that the
existence of nontrivial conservation laws generically leads to ideal
transport properties with infinite Kubo transport coefficients,
and deviation from {\em quantum ergodicity} in general.

In this Letter we present a new and elementary approach
to the construction of IQM 1-dim lattice systems. It is based on the
particular infinite dimensional dynamical Lie algebra (DLA)
generated and represented by the essential dynamical observables
(in our case it is generated by the Ising
Hamiltonian $\sum_j \sigma^x_j \sigma^x_{j+1}$ and the interaction with
the transversal external field $\sum_j \sigma^z_j$) and for
which the `transfer matrix' can be explicitly constructed from
the {\em commutativity condition}. We show that {\em any} element
$H$ of DLA may be considered as a Hamiltonian of IQM system
and construct an analytic DLA valued function
$T(\vec{\lambda})$ of possibly vectorial spectral parameter
$\vec{\lambda}\in{\C}^N$ (for some $N\ge 1$), comuting with $H$,
$[H,T(\vec{\lambda})]\equiv 0$. $T(\vec{\lambda})$ is
a formal analog of {\em logarithm of the transfer matrix}.
The integrals of motion (conserved charges and currents) are
derived as coefficients of Taylor expansion of $T(\vec{\lambda})$
around $\vec{\lambda}=0$.
Therefore we have an infinite-dimensional class of IQM Hamiltonian systems.
Furthermore, {\em real} DLA of self-adjoint observables generates infinite
dimensional unitary dynamical Lie group of even larger class of
integrable quantum many-body propagators of possibly non-Hamiltonian
(non-autonomous, e.g. {\em periodically kicked}) IQM systems.
As an example we work out kicked 1-dim Ising chain
periodically kicked with transversal external field.
Moreover, we explicitly calculate the complete set of eigenstates
and spectra of the conserved charges (including the Hamiltonian).
\\\\
We start with Lie algebra $\frak{U}$ over an infinite spin chain 
spanned by the spatially homogeneous local observables
\begin{equation}
Z_{[s_1 s_2\ldots s_p]} =
\sum_j \sigma^{s_1}_j\sigma^{s_2}_{j+1}\ldots\sigma^{s_p}_{j+p-1},
\label{eq:genob}
\end{equation}
where $\sigma^s_j,s\in\{1=x,2=y,3=z\}$, are the spin variables (Pauli matrices) of
spin $j$ obeying the standard commutation relations
$[\sigma^p_j,\sigma^r_k] = 2 \delta_{jk}\sigma^p_j\sigma^r_k =
2i\delta_{jk}\sum_s\epsilon_{prs}\sigma^s_j$, and $\sigma^0_j = 1$.
The {\em order} of the local observable $A$ is defined as the maximal
number of digits $p$ of some observable (\ref{eq:genob})
in the expansion of $A$ in terms of basis (\ref{eq:genob}).
We are interested in nontrivial infinite dimensional subalgebras of $\frak{U}$
for which the number of elements with order smaller than $p$ grows algebraically
(as a function of $p$) and not exponentially ($\sim 4^p$) as for $\frak{U}$
\cite{footn}.
Indeed we found subalgebra $\frak{S}$, which we call dynamical Lie algebra (DLA)
(essentially generated by $Z_{[3]}$ and $Z_{[11]}$), and spanned by 
two infinite sequences of selfadjoint observables $U_n$ and $V_n$,
\begin{eqnarray}
U_n &=& \left\{ \begin{array}{lll}
     &Z_{[1 (3^{n-1}) 1]}, & n \ge 1, \\
-\!\!&Z_{[3]}, & n = 0, \\
     &Z_{[2 (3^{-n-1}) 2]}, & n \le -1,
\end{array}\right. \label{eq:repr} \\
V_n &=& \left\{ \begin{array}{lll}
     &Z_{[1 (3^{n-1}) 2]}, & n \ge 1, \\
     &Z_{[0]}, & n = 0,\\
-\!\!&Z_{[2 (3^{-n-1}) 1]}, & n \le -1,
\end{array}\right. \nonumber
\end{eqnarray}
for $-\infty < n < \infty$
($(3^n)$ indicates digit $3$ being repeated $n$ times),
which satisfy the following commutation relations
\begin{eqnarray}
\left[ U_{m},U_{n} \right] &=& 2i ( V_{m-n} - V_{n-m} ), \nonumber\\
\left[ V_{m},V_{n} \right] &=& 0, \label{eq:comrel}\\
\left[ U_{m},V_{n} \right] &=& 2i ( U_{m+n} - U_{m-n} ). \nonumber
\end{eqnarray}
The order of observables $U_n$ and $V_n$ is $|n|+1$.
The covering algebra $\frak{U}$ is equiped with the Euclidean
metric associated to the bilinear form (scalar product)
\begin{equation}
(A|B) = \lim_{L\rightarrow\infty}\frac{1}{L 2^{L}}\tr_L (A^\dagger B),
\label{eq:metric}
\end{equation}
($\tr_L$ is a trace for a finite system of size $L$) with respect to which
(\ref{eq:genob}) is an ortho-normal (ON) basis.
Further, $U_n$ and $V_n$ form ON basis of DLA $\frak{S}$ in the same
metric. Note that (\ref{eq:metric}) is {\em invariant}
with respect to {\em adjoint map}, $(\ad A)B = [A,B]$,
namely $((\ad A^\dagger)B|C) = (B|(\ad A)C)$.
\\\\
{\em Conservation laws in general autonomous case:}
Let us assume that the Hamiltonian $H$ and the logarithm of the 
transfer matrix $T$ belong to DLA $\frak{S}$. We write
\begin{eqnarray}
H &=& \sum_{m=-m_-}^{m_+} (h_m U_m + g_m V_m), \label{eq:ham}\\
T &=& \sum_{m=-\infty}^{\infty} (a_m U_m + b_m V_m). \label{eq:transf}
\end{eqnarray}
where the Hamiltonian $H$ has a finite order $M:=\max\{m_+,m_-\}.$
The commutativity relation $[H,T] = 0$ gives the system of
difference equations
\begin{eqnarray}
&&\sum_m h_m (a_{-n+m} - a_{n+m}) = 0,
\label{eq:diffe}\\
&&\sum_m \left[
h_m (b_{n-m} - b_{-n+m}) + g_m(a_{n+m} - a_{n-m})
\right] = 0,
\nonumber
\end{eqnarray}
which can be solved with an ansatz
\begin{eqnarray}
a_n &=& a_+ \lambda^n,\quad b_n = b_+ \lambda^n,\label{eq:ansatz}\\
a_{-n} &=& a_- \lambda^n,\quad b_{-n} = b_- \lambda^n,\nonumber
\end{eqnarray}
for $n \ge 0$. Quite surprisingly, the resulting homogeneous system
\begin{equation}
\pmatrix{h(\lambda) & -h(\lambda^{-1}) & 0 & 0 \cr
       g_a(\lambda) & 0 & h(\lambda^{-1}) & -h(\lambda^{-1}) \cr
       0 & g_a(\lambda) & h(\lambda) & -h(\lambda) \cr}
\pmatrix{a_+ \cr a_- \cr b_+ \cr b_- } = 0
\label{eq:sys}
\end{equation}
has rank 2 for {\em any} value of the {\em spectral parameter} $\lambda$,
where $h(\lambda)$ and $g_a(\lambda) := g(\lambda) - g(\lambda^{-1})$
are the polynomials
$$h(\lambda) = \sum\limits_{m=-m_-}^{m_+} h_m \lambda^m,\quad
  g(\lambda) = \sum\limits_{m=-m_-}^{m_+} g_m \lambda^m.$$
Hence, there are two linearly independent solutions of (\ref{eq:sys})
(up to an arbitrary common prefactor), namely
\begin{eqnarray}
a_+(\lambda) &=& h(\lambda^{-1}),\quad b_+(\lambda) = g(\lambda^{-1}),
\label{eq:sol1} \\
a_-(\lambda) &=& h(\lambda),\qquad b_-(\lambda) = g(\lambda),\nonumber
\end{eqnarray}
and
\begin{eqnarray}
a_+(\lambda)=a_-(\lambda)\equiv 0,\quad b_+(\lambda)=b_-(\lambda)\equiv 1.
\label{eq:sol2}
\end{eqnarray}
(i) Let us first consider the case where the solutions
$a_\pm(\lambda),b_\pm(\lambda)$ are given by (\ref{eq:sol1}).
The global uniform solution (for all $n\in{\Z}$) is given by a linear combination
of $N:=m_{+} + m_{-} + 1$ solutions (\ref{eq:sol1})
\begin{equation}
a_n = \sum_{m=1}^N c_m a_+(\lambda_m)\lambda_m^n,\;\;
b_n = \sum_{m=1}^N c_m b_+(\lambda_m)\lambda_m^n,\;\;
\label{eq:solpos}
\end{equation}
for $n \ge 0$, and
\begin{equation}
a_n = \sum_{m=1}^N c_m a_-(\lambda_m)\lambda_m^{-n},\;\;
b_n = \sum_{m=1}^N c_m b_-(\lambda_m)\lambda_m^{-n},\;\;
\label{eq:solneg}
\end{equation}
for $n \le 0$. 
$N-$tuple of {\em spectral parameters}
$\vec{\lambda}=(\lambda_1\ldots\lambda_N)$ is an arbitrary subset of a
complex unit disk $|\lambda_m|<1$
(in order to ensure convergence of $T$) while the
coefficients $c_m$ are determined by gluing the solutions
(\ref{eq:solpos}) and (\ref{eq:solneg}) on $m_+ + m_- = N-1$ sites around
$n = 0$, giving a homogeneous system of $N-1$ linear equations
\begin{equation}
\sum_{m=1}^N (\lambda_m^n - \lambda_m^{-n}) c_m = 0,\; n=1\ldots N,
\label{eq:systemc}
\end{equation}
with a general (polynomial) solution
\begin{equation}
c_m(\vec{\lambda})=(-1)^m\lambda_m^{N-1}
\prod_{j\le k}^{j,k\neq m}\!(1\!-\!\lambda_j\lambda_k)
\prod_{j<k}^{j,k\neq m}\!(\lambda_j\!-\!\lambda_k)
\label{eq:cm}
\end{equation}
Logarithmic transfer matrix $T(\vec{\lambda})$ is a holomorphic function in 
$\vec{\lambda}$, and the coefficients of its Taylor expansion around 
$\vec{\lambda}=0$ also commute with $H$. After some simple series manipulations 
we easily find an infinite sequence of independent integrals of motion, 
namely the {\em conserved charges} $Q_k,\; k\ge 0$, $[H,Q_k]=0$, 
(note that $Q_0 = 2H$)

\begin{equation}
Q_k = \!\!\sum_{m=-m_-}^{m_+}\! \left[ h_m (U_{k+m}+U_{-k+m})
                         + g_m (V_{k+m}+V_{-k+m})\right].
\label{eq:conslaw1}
\end{equation}
(ii) In the other case, the solutions $a_\pm(\lambda),b_\pm(\lambda)$ 
are given by (\ref{eq:sol2}) and aready solve (\ref{eq:diffe}) 
globally (so $N=1$). The logarithmic transfer matrix is now rather 
trivial, $T(\lambda) = \sum_{m=1}^\infty (V_m + V_{-m}) \lambda^m$,
giving the {\em conserved currents} $C_k,\; k\ge 0$, $[H,C_k] = 0$,
\begin{equation}
C_k = V_{k+1} + V_{-k-1} = Z_{[1 (3^k) 2]} - Z_{[2 (3^k) 1]}.
\label{eq:conslawc}
\end{equation}
$C_0$ is the particle current of the associated spinless fermion model (via
Wigner-Jordan transformation), $C_1$ is the energy current, etc.

It can be easily verified directly that
$ [T_{(i)}(\vec{\lambda}),T_{(i)}(\vec{\mu})] \equiv 0,$
$ [T_{(i)}(\vec{\lambda}),T_{(ii)}(\mu)] \equiv 0,$ and
$ [T_{(ii)}(\lambda),T_{(ii)}(\mu)] \equiv 0.$ Hence all the
conservation laws are in involution
$[ Q_k,Q_l ] = [Q_k,C_l] = [C_k,C_l] = 0.$
For example, for the Ising model in a transversal magnetic field,
$H = J U_1 + h U_0$, one recovers well known conservation laws
$Q_k = J (U_{k+1} + U_{1-k}) + h (U_{k} + U_{-k})$
and $C_k$ (\ref{eq:conslawc}) which required more involved methods in the
literature \cite{G82}.
\\\\
{\em Conservation laws in non-autonomous case,
kicked-Ising model:}
Next we study more general possibly {\em non-autonomous} quantum spin
chains which are propagated by members of a unitary Lie group generated by
DLA $\frak{S}$ which in general {\em cannot} be written in terms of some 
Hamiltonian $H$, as $\exp(-iH)$. For simplicity, we consider {\em periodically 
kicked systems} which correspond to time-dependent Hamiltonian
\begin{equation}
H(t) = H_0 + \delta_p(t) H_1
\end{equation}
where $\delta_p(t)$ is a periodic
delta function of period $1$, and $H_0,H_1 \in \frak{S}$ are the
generators --- the kinetic energy and the potential, respectively.
Using the adjoint representation of
DLA, the (linear) Heisenberg map $U^{\ad}$ of an observable $A\in\frak{S}$
for one time step is factorized as
\begin{equation}
A(t+1) = U^{\rm ad} A(t) = U^{\rm ad}_1 U^{\rm ad}_0 A(t)
\end{equation}
where $U^{\rm ad}_p A = \exp(i\ad H_p) A = \exp(i H_p) A\exp(-i H_p)$,
is the propagation by the kinetic energy and the potential, for
$p=0,1$, respectively.
Transfer matrix is now sought by the invariance condition
\begin{equation}
U^{\ad} T(\vec{\lambda}) = T(\vec{\lambda})
\label{eq:invc}
\end{equation}
in the form
(\ref{eq:transf}). The method of solution is analogous to
(\ref{eq:diffe}-\ref{eq:cm}) whereas the difference equations
for $a_n,b_n$ are now obtained by means of adjoint representation
of propagators which can be derived explicitly by means of eqs.
(\ref{eq:comrel}) and series expansion of exponential function;
say if generated by $U_m$
\begin{eqnarray*}
\exp(i\alpha \ad U_m) U_n &=&
c^2 U_n + s^2 U_{2m-n} + c s (V_{n-m}\!-\!V_{m-n}),\nonumber\\
\exp(i\alpha \ad U_m) V_n &=&
c^2 V_n + s^2 V_{-n} - c s (U_{m+n}\!-\!U_{m-n}),
\end{eqnarray*}
where $c = \cos(2\alpha), s = \sin(2\alpha)$.

Here the general procedure cannot be written as explicitly as
in the autonomous case, so we work out in detail an example of
{\em kicked Ising} (KI) model where the kinetic generator is the
usual 1-dim. Ising Hamiltonian,
$H_0 = J U_1 = \sum_j J \sigma^x_j \sigma^x_{j+1}$,
and the kick potential is the transversal
magnetic field, $H_1 = h U_0 = \sum_j h \sigma^z_j$.
Condition (\ref{eq:invc}) results in
the system of second order difference equations for
$a_n,a_{-n},b_n,b_{-n}$ which is solved thru the ansatz (\ref{eq:ansatz})
giving the solution (again for any $|\lambda|<1$) 
\begin{eqnarray}
a_+(\lambda) &=& s_J c_h + c_J s_h\lambda^{-1},\quad
b_+(\lambda) = s_J s_h(\lambda - \lambda^{-1})/4,\nonumber\\
a_-(\lambda) &=& s_J c_h + c_J s_h\lambda,\qquad
b_-(\lambda) = -b_+(\lambda) \label{eq:solki}
\end{eqnarray}
where $s_J = \sin(2J),c_J = \cos(2J),s_h = \sin(2h),
c_h = \cos(2h)$, and the trivial solution (\ref{eq:sol2}).
In order to obtain the global solution
$(a_n,b_n)$ we again glue together a linear combination of
partial solutions for positive and negative $n$ at the
two points (since the system is of second order), say at $n=0,1$.
We obtain the system (\ref{eq:systemc}) for the three coeficients
$c_m$, $N=3$, depending on a triple of spectral parameters
$\vec{\lambda}=(\lambda_1,\lambda_2,\lambda_3)$
with the solution (\ref{eq:cm}).
Collecting the terms with different powers of $\lambda_m$ 
in the power series expansion of the logarithmic 
transfer matrix $T(\vec{\lambda})$ 
we obtain two infinite sets of conservation laws, namely the charges
$Q_k,\;k\ge 0$, $U^{\ad} Q_k = Q_k$,
\begin{eqnarray}
Q_k &=& c_J s_h (U_{k+1} + U_{-k+1}) + s_J c_h (U_k + U_{-k}) \nonumber \\
    &-& s_J s_h (V_{k+1} + V_{-k+1} - V_{k-1} - V_{-k-1}),
\label{eq:conslaw2}
\end{eqnarray}
and currents (\ref{eq:conslawc}), $C_k,\;k\ge 0,\; U^{\ad} C_k = C_k$.
The conservation laws of KI model (\ref{eq:conslaw2}) are identical to
invariants (\ref{eq:conslaw1}) of an autonomous IQM system with
the Hamiltonian $H_{\rm KI} = Q_0/2 =
c_J s_h U_1 + s_J c_h U_0 - s_J s_h (V_1 - V_{-1})/2.$
Note, however, that the full dynamics are not identical,
$ \exp(i\ad H_{\rm KI}) \neq U^{\ad} $.
\\\\
{\em Structure of DLA and diagonalization of conserved charges:}
Let us now analyze the structure of DLA $\frak{S}$ more carefully.
The current invariants $C_k$ are rather trivial; they span a {\em maximal
ideal} $\frak{I}$ of DLA $\frak{S}$, $[{\frak{S}},{\frak{I}}] = 0$.
The derived (semi-simple) DLA $\frak{S}^\prime 
= [\frak{S},\frak{S}] = (\ad\frak{S})^\infty\frak{S} 
= \frak{S}/\frak{I}$
is spanned by $U_m\pm U_{-m}$ and $V_m - V_{-m}$, for $m \ge 0$,
or in terms of {\em real non-local} Fourier transformed basis
\begin{eqnarray}
J^1(\kappa) &=& \frac{i}{8\pi}
\sum_{m=-\infty}^\infty e^{i\kappa m} (U_m - U_{-m}), \nonumber\\
J^2(\kappa) &=& \frac{i}{8\pi}
\sum_{m=-\infty}^\infty e^{i\kappa m} (V_m - V_{-m}), \label{eq:J123}\\
J^3(\kappa) &=&-\frac{1}{8\pi}
\sum_{m=-\infty}^\infty e^{i\kappa m} (U_m + U_{-m}), \nonumber
\end{eqnarray}
for $0 \le \kappa < \pi$, where the commutation relations read
\begin{equation}
\left[ J^p(\kappa),J^r(\kappa^\prime)\right]
= i\delta(\kappa-\kappa^\prime)\sum_s\epsilon_{prs}J^s(\kappa).
\label{eq:comrel2}
\end{equation}
Therefore, derived DLA is isomorphic to an infinite direct sum
$\frak{S}^\prime \sim \bigoplus_{n=1}^\infty \frak{su}_2$.
It has an infinite rank, Cartan subalgebra is spanned by a continuous
root basis $J^3(\kappa)$ and Chevalley generators are 
$J^\pm(\kappa) = J^1(\kappa)\pm i J^2(\kappa)$.
Now we construct the {\em vacuum state} $\ket{\emptyset}$ 
by the condition $J^{-}(\kappa)\ket{\emptyset}\equiv 0,$
which is is equivalent to 
$(U_m-U_{-m}-iV_m+iV_{-m})\ket{\emptyset}\equiv 0$, and
also $\sigma^-_j\ket{\emptyset}\equiv 0.$ 
Hence the vacuum is the state with {\em all spins down}.
Let us write the Fourier transform of the currents and charges as,
$C(\kappa) = \sum_n \exp(i\kappa m) C_m$
and $Q(\kappa) = \sum_m \exp(i\kappa m) Q_m = Q(-\kappa)$,
respectively. Note that $Q(\kappa) = Q(-\kappa)$ since $Q_m = Q_{-m}$.
Using explicit form (\ref{eq:conslaw1}) we compute
\begin{eqnarray}
Q(\kappa) &=& -8\pi \vec{q}(\kappa)\cdot \vec{J}(\kappa) + 
g_r(\kappa)C(\kappa),\label{eq:FTQ}\\
\vec{q}(\kappa) &=&  (h_i(\kappa),g_i(\kappa),h_r(\kappa)),\nonumber
\end{eqnarray}
where 
$h_r(\kappa)={\rm Re\;} h(\exp(i\kappa)),\; h_i(\kappa)={\rm Im\;} h(\exp(i\kappa))$,
$g_r(\kappa)={\rm Re\;} g(\exp(i\kappa)),\; g_i(\kappa)={\rm Im\;} g(\exp(i\kappa))$.
The structure (\ref{eq:comrel2}) is invariant with respect to arbitrary
local ($\kappa$-dependent) rotation (non-abelian gauge transformation)
of the vector field $\vec{J}(\kappa)$. 
Particularly interesting is the rotation 
${\rm R}(\kappa)$ around axis $\vec{a}(\kappa)$,
$$
\vec{a}(\kappa) = \frac{(g_i(\kappa),-h_i(\kappa),0)}
{\sqrt{g^2_i(\kappa)+h^2_i(\kappa)}},\quad
\vec{a}(\kappa)\cdot\vec{q}(\kappa)\equiv 0,
$$
for an angle $\varphi(\kappa)$,
$$\varphi(\kappa) = 
\arctan\frac{\sqrt{g^2_i(\kappa)+h^2_i(\kappa)}}{h_r(\kappa)},$$
namely,
${\rm R}\vec{r} = 
(\vec{a}\cdot\vec{r})\vec{a} + \frac{\vec{q}\cdot\vec{r}}{|\vec{q}|}\vec{k} +
\frac{(\vec{a}\times\vec{q})\cdot\vec{r}}{|\vec{a}\times\vec{q}|}
\frac{\vec{a}\times\vec{k}}{|\vec{a}\times\vec{k}|},$
where $\vec{k}=(0,0,1)$, which has the property
${\rm R}(\kappa)\vec{q}(\kappa) = |\vec{q}(\kappa)| \vec{k}.$
The unitary transformation of the vector field
$$\vec{W}(\kappa) = {\rm R}(\kappa)\vec{J}(\kappa) = 
\exp(i\!\textstyle{\int_0^\pi}\!\!d\kappa\varphi\vec{a}\cdot\vec{J})
\vec{J}
\exp(-i\!\textstyle{\int_0^\pi}\!\!d\kappa\varphi\vec{a}\cdot\vec{J}),$$ 
makes the the conserved charges $Q(\kappa)$ proportional to the new root 
basis $W_3(\kappa) = \vec{q}(\kappa)\cdot\vec{J}(\kappa)/|\vec{q}(\kappa)|$, 
namely
\begin{equation}
Q(\kappa) = -8\pi |\vec{q}(\kappa)| W^3(\kappa) + g_r(\kappa)C(\kappa).
\label{eq:ch1}
\end{equation}
Using the same rotation we construct a new vacuum state
$\ket{\emptyset}_W$ relative to the field 
$\vec{W}(\kappa)$, $W^-(\kappa)\ket{\emptyset}_W \equiv 0$, namely,
$$
\ket{\emptyset}_W = \exp\left(i\int_0^\pi d\kappa\varphi(\kappa)
\vec{a}(\kappa)\cdot\vec{J}(\kappa)\right)\ket{\emptyset}. 
$$
Let us now discretize the momentum $\kappa$ to $L$ bins what corresponds 
to (but is not identical to) a finite chain of $L$ spins, and define
$$W^p_k:=\int_{\pi(k-1)/L}^{\pi k/L} \!\!d\kappa W^p(\kappa), 
\quad 1\le k \le L.$$
Then we have $[W^p_k,W^r_l]=i\delta_{kl}\sum_s \epsilon_{prs}W^s_k$.
The eigenstates of conserved charges $Q(\kappa)$ (and of 
root basis $W^3_k$ since $[Q(\kappa),W^3(\kappa^\prime)]\equiv 0$) 
can be labeled by $L$ binary quantum numbers 
$c_k\in\{0,1\},k=1\ldots L$, and are constructed by means of 
creation operators
\begin{equation}
\ket{c_k} = \prod_{1\le k\le L}^{c_k=1} W^{+}_k\ket{\emptyset}_W,
\label{eq:eigsW}
\end{equation}
with $W^3_l\ket{c_k} = (c_l-1/2)\ket{c_k}$. 
Hence, for smooth $|\vec{q}(\kappa)|$ and large $L$, all
the charges (\ref{eq:ch1}) are {\em diagonal} in the eigenbasis
(\ref{eq:eigsW}).
Of course, the {\em eigenvalues} are finite (for the infinite system
$L=\infty$) only for the charge {\em densities} 
$Q^\prime_m = \lim_{L\rightarrow\infty}(1/L)Q_m|_L$
and not for the {\em extensive} charges $Q_m$. 
The eigenvalues of invariant densities are computed by taking the 
limit $L\rightarrow\infty$ and the inverse cosine transform
$$
Q^\prime_m\ket{c(\kappa)}
= -\frac{8}{\pi}\int_0^\pi d\kappa^\prime\!\!\cos(\kappa^\prime m)
|\vec{q}(\kappa^\prime)|\left[c(\kappa^\prime)-\frac{1}{2}\right]\ket{c(\kappa)}.
$$
Label $c(\kappa)$ is an arbitrary (measurable) {\em index function}
$c : [0,\pi)\rightarrow \{0,1\},$ 
(for finite $L$, $c(\kappa) = c_k, \pi(k-1)/L\le \kappa < \pi k/L$) 
and $\ket{c(\kappa)}$ is the corresponding eigenstate
which should be properly defined by some limiting procedure 
$L\rightarrow\infty$ of (\ref{eq:eigsW}). 
It seems that the spectrum of $Q^\prime_m$ is purely continuous.
On the other hand, for the currents we have $C_m\ket{c(\kappa)}\equiv 0$,
since $[C(\kappa),\vec{W}_l]\equiv 0$.
Having such a transparent structure (\ref{eq:comrel2}-\ref{eq:eigsW})
it should be an easy task to compute physically interesting 
{\em correlation functions}.
\\\\
In this Letter we have introduced an {\em infinitely dimensional
space} of completely integrable quantum many-body systems (spin-chains or
chains of spinless fermions),
the so called Dynamical Lie Algebra, as opposed to few-parameter
families of completely integrable quantum many body systems known so far
in the literature.
The model is an infinitely dimensional extension of Ising
model in transversal field (equivalent to XY-model \cite{G82} and to
1-dimensional free fermion theory).
For every element of the algebra being interpreted as a Hamiltonian,
or any propagator from the associated unitary Lie group being generated
by a finite number of elements of the algebra (e.g. Ising model periodically
kicked by transversal magnetic field),
we construct two infinite sets of quantum invariants of motion, the conserved
charges and the conserved currents.
Is is shown heuristically how to diagonalize these conservation laws.
Explicit expressions of the conserved charges are quite simple
(much simpler than in general Heisenberg (XYZ) or Hubbard model\cite{GM95},
for example) though nontrivial.

Financial support by the Ministry of Science and Technology of R Slovenia is 
gratefully acknowledged.

\end{document}